\documentstyle[aps,epsfig,preprint]{revtex}

\begin{document}
\draft
\title{Thermal relaxation of field-induced irreversible ferromagnetic phase in Pr-doped
manganites }
\author{Joonghoe Dho and N. H. Hur\cite{NHHur}}
\address{Center for CMR materials, Korea Research Institute of Standards and Science,
Yuseong P.O.Box 102, Daejon 305-600, Korea}
\date{\today}
\maketitle

\begin{abstract}

We have investigated the magnetic and transport properties of
Pr$_{0.65}$Ca$_{0.35}$MnO$_3$ and Pr$_{0.82}$Na$_{0.18}$MnO$_3$
with almost same doping level under various magnetic fields. Upon
applying a field of 4 T, both compounds undergo a metamagnetic
transition from antiferromagnetic (AFM) to ferromagnetic (FM) in
the low temperature region. The field-induced FM state is very
stable for at least 10$^5$ s even after removing the field. This
irreversible FM behavior is different from the spontaneous FM
ordering as well as the FM state in a reversible metamagnetic
transition. We discuss this unusual FM phenomenon found in the
Pr-doped manganites with the AFM ground state in terms of the
reorientation of Mn spins pinned by localized Pr moments.

\end{abstract}
\pacs{PACS number : 75.30.Vn, 75.30.Kz, 75.60.Nt}

%\begin{multicols}{2}
\narrowtext

\section{Introduction}

Over the last few years, a large amount of research has been
undertaken towards understanding the physical properties of doped
perovskite manganites, mainly because they exhibit a novel magneto
transport property known as colossal magneto resistance
(CMR)\cite{zener,tokura,coey}. Among them, Pr-doped manganites
Pr$_{1-x}$A$_x$MnO$_3$ (A = Ca, Sr, Na) have been the subjects of
several studies due to their diverse physical and structural
properties with doping concentration at different
temperatures\cite{yoshizawa,tomioka,moritomo,radaelli,deac,jirak,hejtmanek,satoh}.
In the case of a 30\% Ca doped sample, it undergoes a charge
ordering (CO) transition near 220 K ($T_{CO}$), an AFM transition
about 150 K ($T_N$), and a FM transition around 100 K ($T_C$).
Below the AFM transition temperature, interestingly, application
of a high magnetic field leads to the FM and metallic state.
Although the nature of the induced FM and metallic phase has not
been completely understood yet, it has been suggested that the
induced state might be due to the development of FM clusters in
the CO and AFM matrix after the application of a magnetic field.
This implies that the field-induced state has its origin in a
change of the electronic state of the Mn ions and is not directly
related to the Pr moments. Neutron diffraction study on
Pr$_{0.7}$Ca$_{0.3}$MnO$_3$ by Cox et al, however, revealed the
presence of ordered Pr 4f moments below 60 K\cite{cox}. Very
recently, Dho et al also reported evidence of long range ordering
of the Pr spins in Pr$_{0.8}$Ca$_{0.2}$MnO$_3$, implying that the
field induced FM and metallic state might be associated with the
development of the ordered Pr moments\cite{dho1}.

In this paper, we present various magnetization and relaxation
data for two Pr-doped manganites Pr$_{0.65}$Ca$_{0.35}$MnO$_3$ and
Pr$_{0.82}$Na$_{0.18}$MnO$_3$ with almost identical doping
concentration. Our results clearly reveal that the field induced
FM moment is present in both materials. Furthermore, this FM phase
is remarkably sustained for 10$^5$ s even after the field is
removed. It is the main purpose of this paper to show that the
field induced irreversible FM state is different from the
metamagnetic transition as well as the typical spontaneous FM
moment. In addition, we discuss the origin of this irreversible FM
state, which might be due to the ordered Pr moments that are able
to control the direction of neighboring Mn spins.

\section{Experimentals}

Several single crystals of Pr$_{0.65}$Ca$_{0.35}$MnO$_3$,
corresponding to 35 \% doping of holes onto the Mn sites, were
grown by a floating zone method using an infrared image furnace.
Many attempts to grow single crystal of
Pr$_{0.82}$Na$_{0.18}$MnO$_3$ with the same floating zone method
were unsuccessful mainly because the sodium is rapidly evaporated
upon melting. For the Na-doped compound, ploycrystalline samples
synthesized by conventional solid state reaction were thus used.
X-ray powder diffraction data of the crushed crystalline and
polycrystalline samples at room temperature show them to be a
single phase, in which both samples adopt an orthorhombic Pnma
symmetry. The magnetization data were collected on a
superconducting quantum interference device (SQUID) magnetometer,
and the resistivity data were collected using four-point method.

\section{Results}
\subsection{Irreversible field-induced phase transition in Pr$_{0.65}$Ca$_{0.35}$MnO$_3$}

The temperature-dependent dc magnetization and resistivity data of
Pr$_{0.65}$Ca$_{0.35}$MnO$_3$, taken under both zero-field-cooled
(ZFC) and field-cooled (FC) conditions, are presented in Figure 1.
The magnetization curves below 3 T did not reveal any discernible
change indicative of a FM transition. Instead, two broad humps are
seen at 160 and 230 K, which correspond to AFM and charge ordering
transitions, respectively. Above 4 T, a clear field-induced
transition is observed near 100 K, indicating that a metamagnetic
transition from AFM to FM takes place by the application of a
magnetic field. In addition, the M($T$) curves display a large
splitting between the FC and ZFC curves below 25 K, indicating the
existence of spin-glass-like state in the low temperature region.
The temperature dependent resistivity data shown in the bottom
panel are well correlated with the M($T$) curves discussed above.
The resistivity curve exhibits a steep increase near $T_{CO}$ and
an insulator-to-metal (I-M) transition occurs in the presence of a
high magnetic field near the field-induced FM transition
temperature. A clear splitting between the ZFC and FC resistivity
data is observed below 25 K, which is consistent with the
splitting in the M($T$) data.

To further understand the nature of the field-induced FM
transition, we have explored the magnetization as a function of
magnetic field at selected temperatures (Figure 2). Each panel
includes an initial magnetization curve and a hysteresis loop
between -5.5 T and +5.5 T. The M(H) curves measured at 50 K
exhibit a distinctive metamagnetic transition from AFM to FM near
4 T. When the magnetic field is released, the field-induced FM
state is returned to an initial AFM state near 2.5 T. As can be
seen in the middle panel of Figure 2, however, the M(H) data at 20
K show that the initial field sweep curve yields an irreversible
change. In particular, the induced FM moment partially remains
even after field quenching. Furthermore, at 5 K the hysteresis
loop exhibits little evidence of an AFM signature although the
initial magnetization curve shows a metamagnetic transition. The
hysteresis curve indicates a soft FM character without hysteretic
energy loss. This FM state is thus apparent in the low temperature
region, implying that the field-induced FM state becomes stable
after the measurement of the initial magnetization curve. An
important feature is that this irreversible field-induced phase
transition is different from a reversible metamagnetic transition
typically observed in the AFM material.

Figure 3 shows the resistivity curves for
Pr$_{0.65}$Ca$_{0.35}$MnO$_3$ as a function of magnetic field
measured at 5 and 50 K. At 50 K, the resistivity suddenly drops
near 4 T, indicating that the charge-ordered insulating phase is
melted into the FM metallic state by the external field. Upon
reducing the magnetic field, the metallic phase is gradually
changed into the charge-ordered ground state. The resistivity data
between -5.5 T and 5.5 T also show a hysteretic behavior as found
in the magnetization data. However, the resistivity curve at 5 K
shows a distinctive I-M transition about 4 T and the field-induced
metallic state is sustained even after removing the magnetic
field. All these transport behaviors are well consistent with the
magnetization data discussed above.

The stability of the field-induced FM moments in the low
temperature region was examined by measuring the long-time
relaxation of the dc magnetization. The sample was cooled with a
field of 5 T down to the measured temperature (5 T) and the time
dependent magnetization data were then collected immediately after
setting the magnetic field at 0.3 T. Figure 4 shows the M(t)
curves as a function of time. An important feature is that the
magnetization at 5 K was roughly constant for 10$^5$ s without any
noticeable change. With increasing temperature, however, the
magnetization gradually relaxes to the value of the AFM state.
These results suggest that the field-induced FM moments are very
stable at 5 K but are gradually destroyed at higher temperature.

The temperature dependence of the field-annealed FM moment was
also investigated in order to compare with the ZFC magnetization
curve shown in Figure 1. The sample was cooled down to 5 K in 5 T,
followed by removing the magnetic field for 10$^3$ s. The
magnetization data (filled symbol) were then collected immediately
after setting a field of 0.3 T. As can be seen in the M($T$) curve
given in Figure 5, a distinctive magnetic transition from FM to
AFM is observed near 25 K. The ac susceptibility data obtained in
zero field also exhibits almost identical temperature dependent
behavior. On the other hand, the ZFC magnetization curve (open
symbol) obtained at the same magnetic field (0.3 T) does not show
any tangible FM transition over the entire temperature range. The
magnetic transition from FM to AFM near 25 K is also accompanied
by a metal-insulator transition. This clearly suggests that the FM
transition below 25 K results from the field annealing.

\subsection{Irreversible field-induced phase transition in Pr$_{0.82}$Na$_{0.18}$MnO$_3$}

In order to probe the role of the Pr concentration, we have also
investigated the magnetization and transport properties of
Pr$_{0.82}$Na$_{0.18}$MnO$_3$. Since the Na-doped compound has
almost the same doping level as well as the same average A-site
ionic size by comparison with those of
Pr$_{0.65}$Ca$_{0.35}$MnO$_3$, its charge ordering and AFM
transitions occur near the temperatures observed in
Pr$_{0.65}$Ca$_{0.35}$MnO$_3$\cite{coey,hejtmanek}. An important
difference, however, is that the Pr concentration is much larger
in Pr$_{0.82}$Na$_{0.18}$MnO$_3$ than in
Pr$_{0.65}$Ca$_{0.35}$MnO$_3$. This would enable to see the
pronounced Pr effect in the Na-doped sample that is crucial to
elucidate the origin of the irreversible field induced
ferromagnetism.

Figure 6 displays the magnetization curves of
Pr$_{0.82}$Na$_{0.18}$MnO$_3$ as a function of magnetic field at
selected temperatures. The hysteresis loops between +5.5 T and
-5.5 T (filled circle) were recorded just after the initial
measurement from 0 T to +5.5 T (open circle). Above 100 K, the
M(H) curves exhibit a metamagnetic transition from AFM to FM with
increasing the magnetic field. The critical magnetic field
decreases with decreasing the temperature. In addition, the
field-induced FM state is returned to an initial AFM state upon
removal of the magnetic field. At 80 K, however, the M(H) data
show that the initial field sweep curve yields an irreversible
change similar to that of Pr$_{0.65}$Ca$_{0.35}$MnO$_3$ measured
at 20 K. In particular, the induced FM moment is sustained even
after field quenching. When the magnetic field is swept from 0 T
to -5.5 T, the M(H) curve displays both FM and AFM signatures at
low and high field regions, respectively. Below 60 K, the initial
magnetization curve shows a metamagnetic transition but other M(H)
data clearly exhibit a typical soft FM feature. It is worthwhile
to note that the field-induced FM state is found in much broader
region in Pr$_{0.82}$Na$_{0.18}$MnO$_3$ compared with that of
Pr$_{0.65}$Ca$_{0.35}$MnO$_3$, which is due to the large Pr
concentration in the Na-doped compound.

The temperature dependence of the field-annealed FM moment of
Pr$_{0.82}$Na$_{0.18}$MnO$_3$ was also investigated, which is
given in Figure 7. The sample was cooled down to 5 K in the
presence of a magnetic field of 5 T, followed by quenching the
field for 10$^3$ s. The temperature dependent magnetization data
(filled squares) were then collected immediately after setting a
field of 0.3 T. As can be seen in the M(T) curve, a distinctive FM
transition is observed near 90 K. On the other hand, the
magnetization data without the field annealing (open squares) show
little indication of a FM transition. These magnetic data also
coincide with the resistivity curves. The field-annealed sample
shows a distinctive I-M transition near the FM transition
temperature while the sample without the field annealing is
insulating. An interesting feature to be noted is that the
irreversible field-induced FM phase is also discovered in other
Pr-doped manganites such as Pr$_{0.7}$Ca$_{0.3}$MnO$_3$,
Pr$_{0.67}$Ca$_{0.33}$MnO$_3$, and
Pr$_{0.65}$Ca$_{0.35}$MnO$_3$\cite{dho2}. The $T_C$ of the
field-induced FM phase monotonously decreases from 60 K to 25 K
with decreasing the Pr concentration. More interestingly, no
irreversible field-induced FM transition behavior is found in
Pr$_{0.6}$Ca$_{0.4}$MnO$_3$ with the low Pr content, clearly
suggesting that the field-induced FM exchange coupling is
dependent on the Pr concentration.

\section{Discussions}

It is worthwhile to address that the FM moment of Pr-doped
manganites, produced by field annealing is different from a
spontaneous magnetic moment found in a typical FM material as well
as in the reversible metamagnetic transition. A distinctive
feature in the field-annealed FM phase is that the ferromagnetism
is induced by field annealing and survives even after field
quenching. Now the question is how this irreversible field-induced
FM phase in Pr$_{0.65}$Ca$_{0.35}$MnO$_3$ and
Pr$_{0.82}$Na$_{0.18}$MnO$_3$ is stabilized in the CO and AFM
matrix. To understand this unusual phenomenon, we employ a
schematic energy diagram (Figure 8) that includes two local minima
corresponding to the AFM and FM states. In zero magnetic field,
the ground state of the two compounds is the pseudo CE-type CO and
AFM state. As given in Figure 8(a), the FM state is not stable at
this stage. In the presence of a high magnetic field, however, the
FM phase is stabilized by means of the Zeemann energy term
(-M$\cdot$H), which is depicted in Figure 8(b). In the ordinary
AFM material, the field-induced FM state immediately goes back to
the AFM ground state when the field is removed. The field-induced
FM phase in the Pr-doped manganites to be discussed is
irreversible and stable even after quenching the field. This can
be described by the introduction of pinning potential barrier (U),
given in Figure 8(c). The only way to destabilize the FM state is
the increase of the thermal fluctuation in the high temperature
region. Above a critical temperature ($T_0$), the thermal
fluctuation abruptly induces a transition from FM to AFM, as seen
in Figure 5 and 7. Accordingly, we can simply estimate that the
pinning potential barrier is $\sim k_B T_0$, where $k_B$ is the
Boltzmann constant, which is corresponding to about 2-8 meV in
Pr-doped manganite samples. Our simple energy diagram well
explains the reversible and irreversible FM states and also agrees
with the M(H) curves given in Figure 2 and 6.

Another important issue to be addressed is that spin reorientation
is the most likely cause of the irreversible field-induced FM
state found in the Pr-doped
manganites\cite{fridrikh,lumsden,tsujii,lees}. In zero magnetic
field, FM exchange coupling between the Pr 4f and Mn 3d spins is
negligible because the Pr spins have no net moments in the AFM
matrix. With an increase in the magnetic field, however, the Mn
spins undergo a metamagnetic transition and simultaneously the
localized Pr spins change to an ordered state. Consequently, the
FM exchange coupling between the Pr and Mn spins is operative,
making the field-induced FM state of Mn spins stable. The origin
of this unique FM state thus appears to be associated with the
coupled interaction between trivalent Pr ions and mixed valent Mn
ions. Probably the ordered 4f Pr spin might pin the itinerant Mn
spins. Namely they control the spin direction of Mn 3d electrons
through the FM exchange coupling, which eventually leads to the
development of the ferromagnetism. Accordingly, this field
annealed FM state is pinned by the ordered 4f spins, which makes
the FM phase stable. It is conceivable that the Pr 4f electrons
may act as a pinning center for pinning itinerant charge carriers.
This anomalous irreversible switching of the FM interaction is
strongly dependent upon the order-disorder transition of the
localized Pr spins.

Finally, it is worthy to mention that this field-induced
irreversible FM phase is not easily observed in the Nd-doped
samples Nd$_{1-x}$Ca$_x$MnO$_3$ although sizes and magnetic
moments of the Pr and Nd ions are similar\cite{tokunaga,fan}. One
likely cause of this difference is ascribed to the different
electronic states between the Pr and Nd ions. Unlike the Nd ion
that has predominantly a 3+ state, the Pr ion can have a 4+ state
as well as a 3+ state. This mixed valent nature appears to induce
the strong FM exchange coupling in the Pr-doped manganite
efficiently. Another plausible cause is due to subtle size
difference between the Nd and Pr ions. The smaller Nd ion leads to
slightly larger lattice distortion, which makes the FM exchange
coupling between the Nd 4f and Mn 3d spins coupling difficult.

\section{Conclusions}

We have discovered a new type of ferromagnetism in Pr-doped
manganites. The observed ferromagnetism is quite unusual in that
the field annealing yields an irreversible field-induced FM state.
This irreversible field-induced FM state is due to an irreversible
switching of the FM interaction between the Pr and Mn spins by the
magnetic field. If one calls a spontaneous magnetic moment
typically observed in a FM material {\it type I ferromagnetism},
then this field-induced ferromagnetism might be dubbed {\it type
II ferromagnetism}. We have adopted this terminology from type II
superconductor. This is because although they are intrinsically
different, they have some similar features, particularly in a
mixed state. The FM moment induced by the ordered Pr spins that
pin the itinerant charge carriers will play an important role in
determining the physical properties of
Pr$_{0.65}$Ca$_{0.35}$MnO$_3$ and Pr$_{0.82}$Na$_{0.18}$MnO$_3$.
Indeed, we have found that the order-disorder transition of
localized Pr spins is strongly correlated with a metal-insulator
transition in the field-annealed sample of
Pr$_{0.65}$Ca$_{0.35}$MnO$_3$.

\section{Acknowledgement}

The Creative Research Initiative Program supported this work.

\newpage

\begin{figure}
\epsfig{file=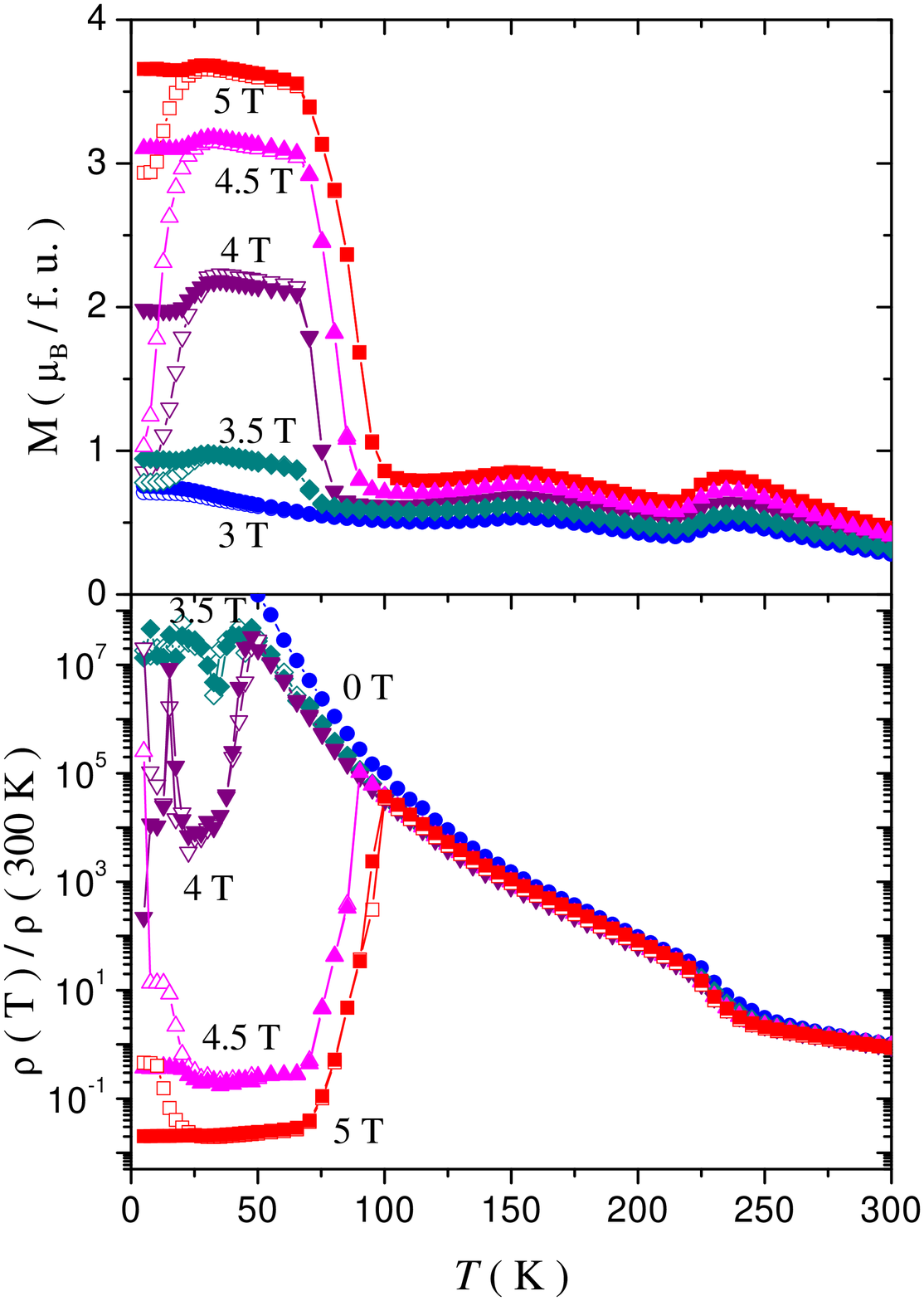,width=12cm} \caption{Temperature dependent
magnetization of Pr$_{0.65}$Ca$_{0.35}$MnO$_3$ taken in various
magnetic fields during both the zero-field-cooled (ZFC, open
symbol) and field-cooled (FC, filled symbol) runs.  }
\end{figure}

\newpage

\begin{figure}
\epsfig{file=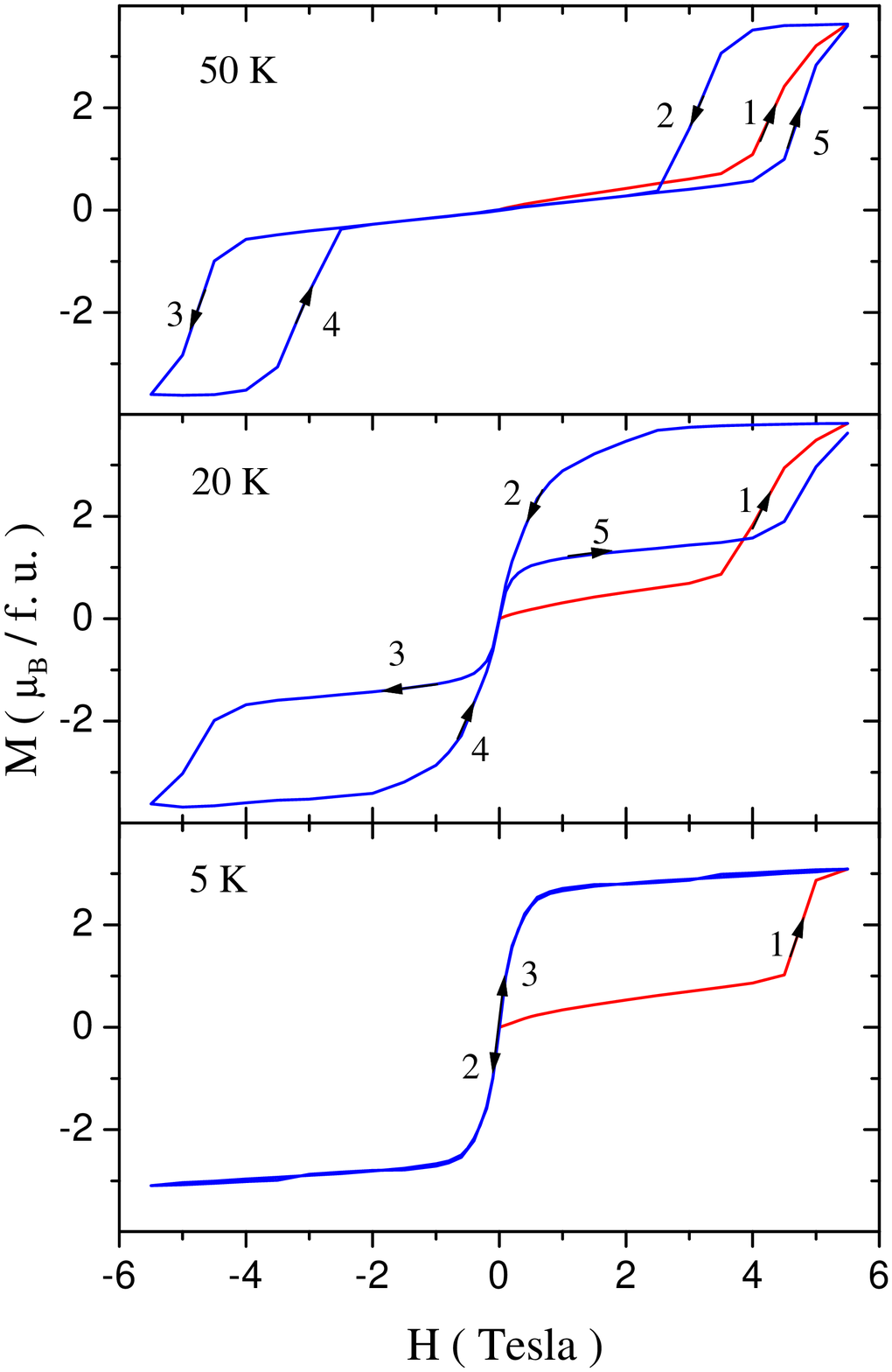,width=12cm} \caption{Field dependence of
magnetization measured at selected temperatures for
Pr$_{0.65}$Ca$_{0.35}$MnO$_3$. Measurements have been carried out
in numerical sequence. }
\end{figure}

\newpage

\begin{figure}
\epsfig{file=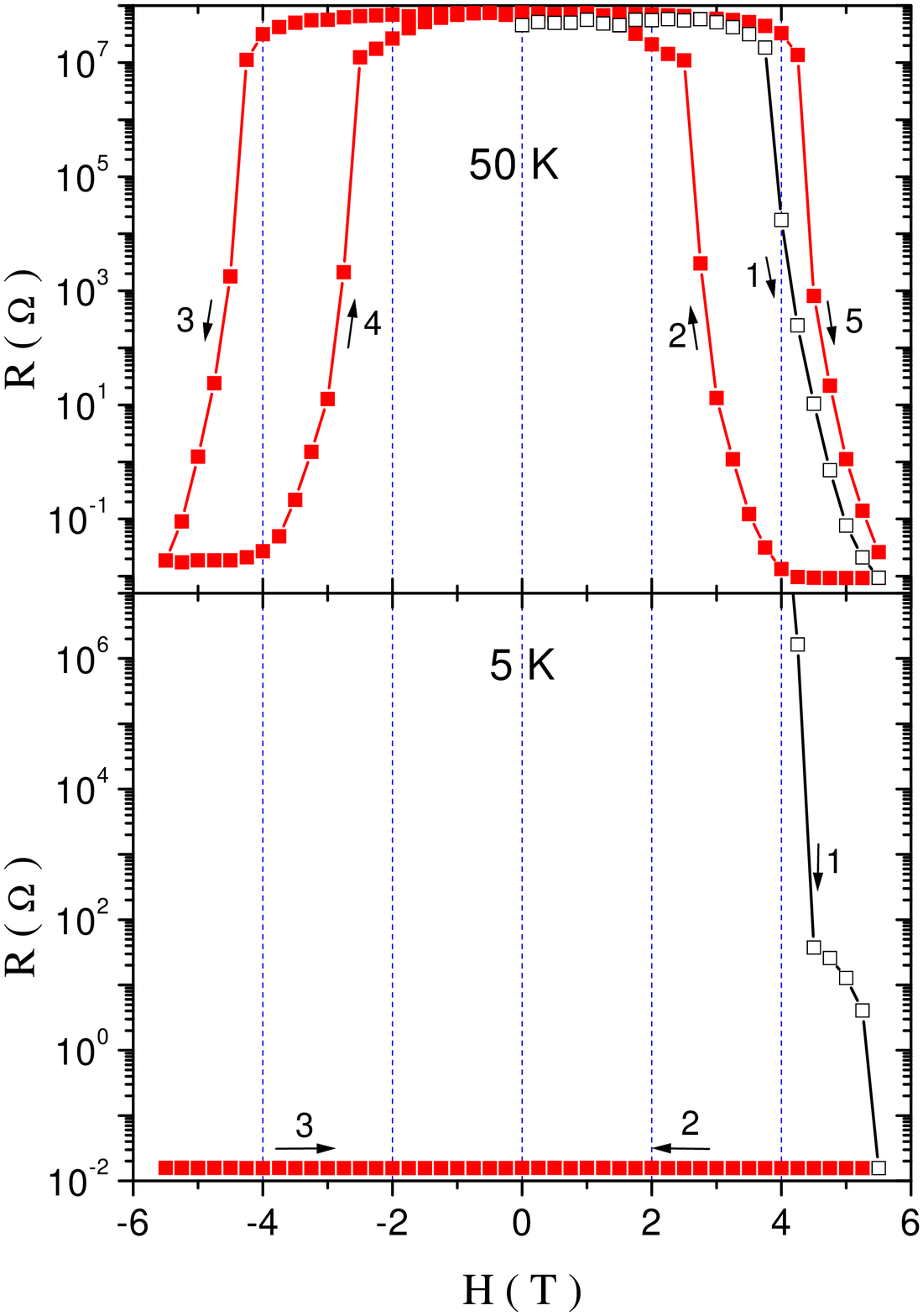,width=12cm} \caption{Field dependence of
resistivity measured at selected temperatures for
Pr$_{0.65}$Ca$_{0.35}$MnO$_3$. Measurements have been carried out
in numerical sequence. }
\end{figure}

\newpage

\begin{figure}
\epsfig{file=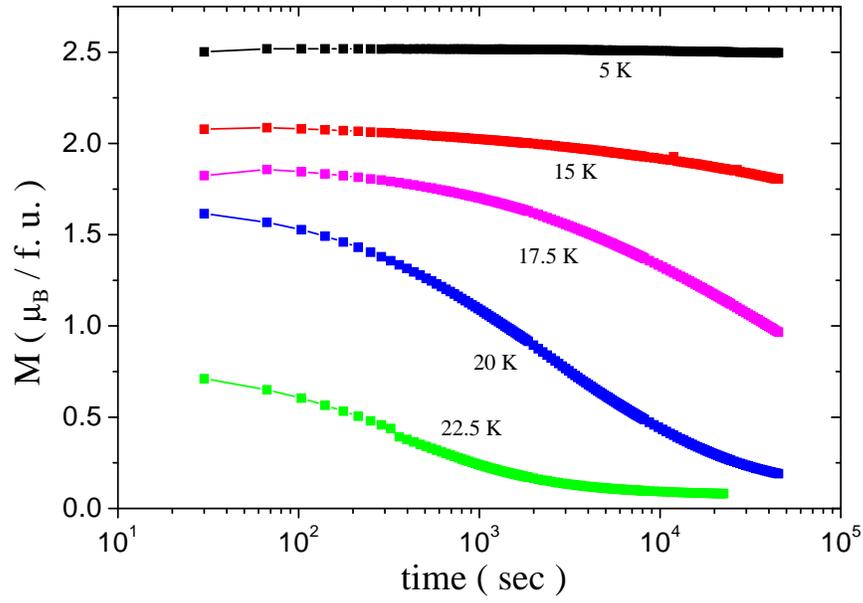,width=12cm} \caption{Time dependence of
magnetization for Pr$_{0.65}$Ca$_{0.35}$MnO$_3$ at several
temperatures. The sample was initially cooled from 300 K down to
the measured temperature under 5 T, followed by measuring the time
dependent magnetization immediately after setting the field to 0.3
T.}
\end{figure}

\newpage

\begin{figure}
\epsfig{file=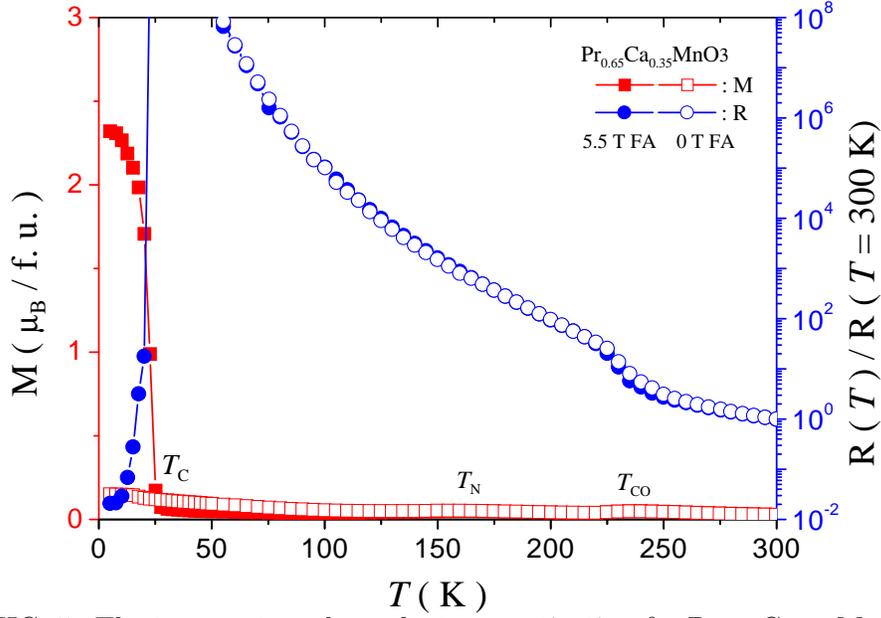,width=12cm} \caption{ The temperature
dependent magnetization for Pr$_{0.65}$Ca$_{0.35}$MnO$_3$ measured
after magnetic field-annealing (FA). The sample was cooled down to
5 K at 0 T (open symbol) and 5 T (filled symbol) in respective
runs, followed by removing the magnetic field for 10$^3$ s. The
magnetization and resistivity measurements were performed in 0.3 T
and 0 T, respectively. All the data were collected in warming
cycle. }
\end{figure}

\newpage

\begin{figure}
\epsfig{file=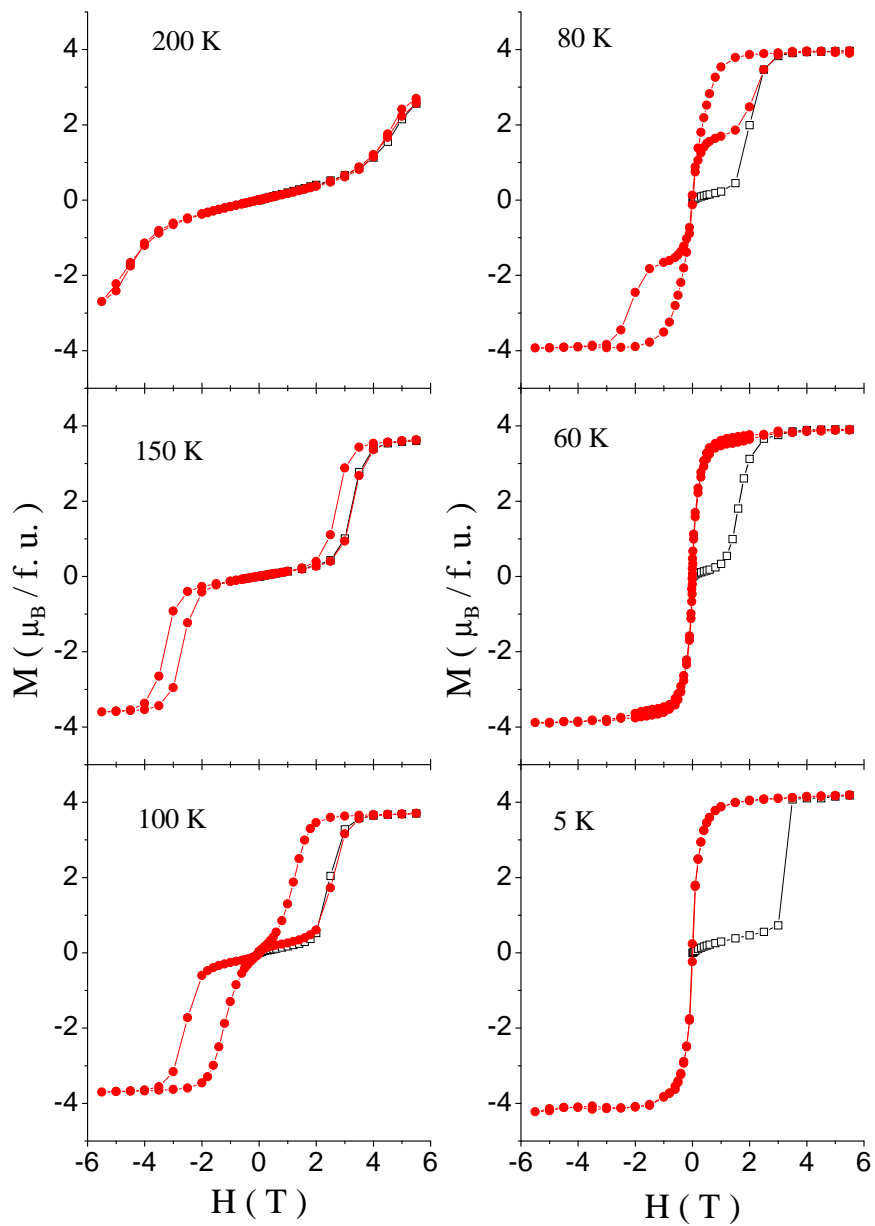,width=12cm} \caption{The field dependent
magnetization curves measured at selected temperatures for
Pr$_{0.82}$Na$_{0.18}$MnO$_3$. The open squares represent the
initial magnetization data and the filled squares denote the
subsequent hysteresis loop. }
\end{figure}

\newpage

\begin{figure}
\epsfig{file=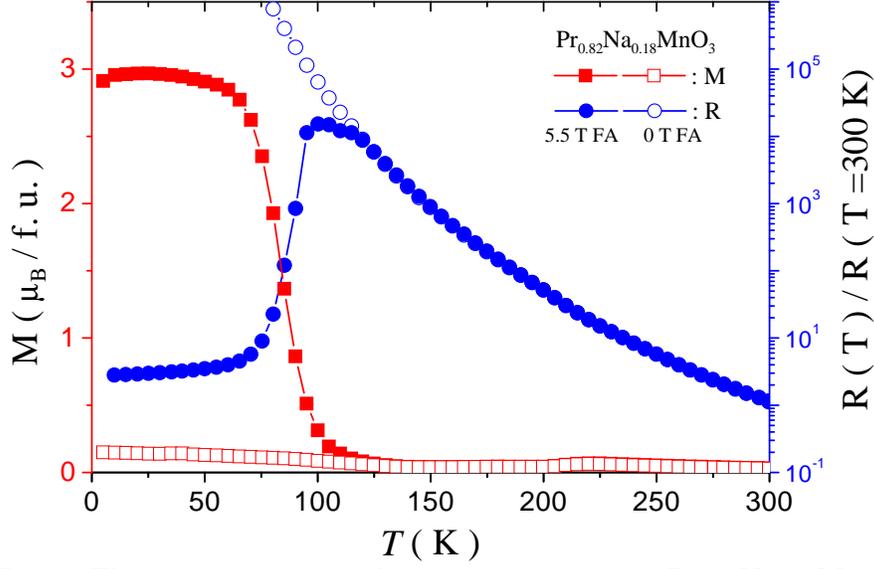,width=12cm} \caption{The magnetization and
resistivity curves for Pr$_{0.82}$Na$_{0.18}$MnO$_3$ as a function
of temperature. The sample was cooled down to 5 K at 0 T (open
symbols) and 5 T (filled symbols), followed by removing the
magnetic field for 10$^3$ s. The magnetization and resistivity
measurements were performed in 0.3 T and 0 T, respectively. All
the data were collected in the warming cycles.  }
\end{figure}

\newpage

\begin{figure}
\epsfig{file=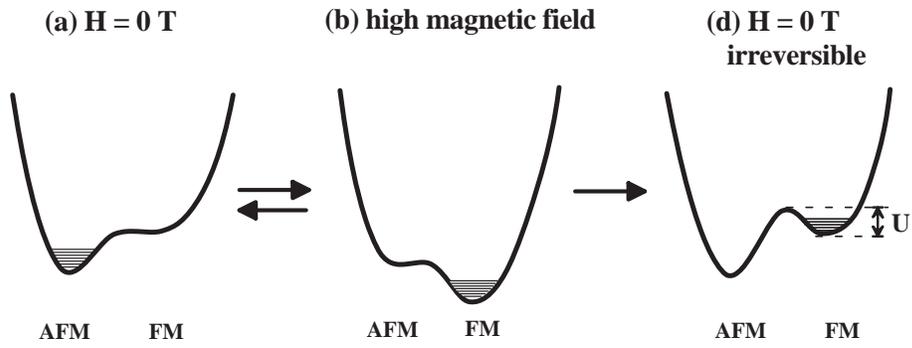,width=12cm} \caption{Schematic energy
diagrams of FM and AFM states with and without the external
magnetic field, in which two local minima represent potential
energy states of AFM and FM phases.}
\end{figure}

%\end{multicols}
\end{document}